\documentclass[a4paper, 11pt]{article}
\usepackage{vmargin}
\setmarginsrb{2.0cm}{3.0cm}{2.0cm}{1.0cm}{0.0cm}{0.0cm}{0.0cm}{1.0cm}
\usepackage[french,english]{babel}
\usepackage{graphicx, float, setspace,amssymb,amsmath}
\usepackage{color}

\begin{document}
\selectlanguage{english}
 
\title{Magnetic and structural properties of nanocrystalline PrCo$_3$.
}

\author {K Younsi$^1$, V Russier$^1$, L Bessais$^1$, J-C Crivello$^1$ \\
$^1$ ICMPE, UMR 7182 CNRS-Universit\'e Paris Est, \\
2-8 Rue Henri Dunant F-94320 Thiais, France. \\
russier@icmpe.cnrs.fr}
\date{}
\maketitle

\begin{singlespace}
\abstract{

The structure and magnetic properties of nanocrystalline PrCo$_3$ obtained from
high energy milling technique are investigated by X-ray diffraction,
Curie temperature determination and magnetic properties measurements are reported.
The as-milled samples have been annealed in a temperature range of 1023~K to
1273~K for 30~mn to optimize the extrinsic properties.
The Curie temperature is 349\,K and coercive fields of 55\,kOe at 10\,K and 12\,kOe at
293\,K are obtained on the samples annealed at 1023\,K. A simulation of the
magnetic properties in the framework of micromagnetism has been performed in order to
investigate the influence of the nanoscale structure. A composite model with hard
crystallites embedded in an amorphous matrix, corresponding to the as-milled
material, leads to satisfying agreement with the experimental magnetization curve.
[ K. Younsi, V. Russier and L. Bessais, J. Appl. Phys. {\bf 107}, 083916 (2010)].
The microscopic scale will also be considered from DFT based calculations of the electronic structure of $R$Co$_x$ compounds, where $R$ = (Y, Pr) and $x$ = 2,3 and 5.
}
\end{singlespace}

\section{Introduction}
The permanent magnetic materials with high performance require appropriate intrinsic hard magnetic properties {\it i.e} 
a high uniaxial magnetocrystalline anisotropy field, a high Curie temperature and a high saturation magnetization.
Among nanocrystalline magnets based on rare earth and transition metal compounds
the $R$Co$_3$ compounds ($R$ as a rare earth)
have been the subject of many studies during the last decades.
These ones crystallize
in rhombohedral PuNi$_3$-type (space group: $R\bar3m$). The unit cell contains two non-equivalent crystallographic sites for  $R$~ions, $3a$ and $6c$, and three sites for Co: $3b$, $6c$ and $18h$. $R$Co$_3$ alloys present excellent magnets properties such as large magnetocrystalline anisotropy and important saturation magnetization. These properties are resulting from to combination of the $3d$ itinerant magnetism of the Co sublattice and $4f$ localized magnetism of the sublattice.

In this study,
we report the elaboration of nanocrystalline PrCo$_3$ by high-energy milling, and their structural and magnetic properties in amorphous state and after recrystallization. 
Hard magnetic properties of the as milled PrCo$_3$ have been improved by a controlled nanocrystallization, which optimize the 
grain size in order to obtain a high coercivity.

In addition to the experiments,
results from simulation and microscopic calculations are presented. We have proposed a simple model for the system after annealing based on a
lattice of crystallites embedded in an amorphous matrix. The simplest realization of such a model is considered here, in order to bring out at the qualitative level the microstructure effect on the magnetization curve. This model is treated in the framework of micromagnetism, {\it i.e.} through a continuous medium type of approach, and the calculations are performed by using the \mbox{MAGPAR} \cite{magpar} code based on the finite elements method.
We also present the results of {\it ab initio} band structure calculations and magnetic properties based on the spin polarized local approximation of the density functional theory (DFT) of $R$Co$_x$, where $R$=(Y,Pr) and $x$=2,3 and 5 intermetallic compounds. We will focus on a quantitative analyse of the localized $4f$~states and itinerant $3d$ magnetism.

\section{Experimental studies}
A polycrystalline PrCo$_3$ was prepared by melting high-purity starting elements ($>99.9\%$) in an induction furnace under
an atmosphere of high purity Ar. The ingot was remelted five times to ensure homogeneity. These ingots were used as pre-alloys to manufacture samples by high-energy milling. Technique and conditions of milling has been described in \cite{yrb10}.
After milling, the powder mixtures was wrapped in tantalum foil and sealed in silica tubes under a vacuum of  10$^{-6}$\,Torr, then annealed for 30\,min at temperature between
1023~K and 1323~K
followed by quenching in water.
X-ray diffraction (XRD) patterns of as-cast, as-milled and annealed powders were carried out with Cu-K$\alpha $ radiation source on a Bruker diffractometer. The unit-cell parameters were measured with Si as the standard  ($a$= 5.4308\,\AA) leading to a unit-cell parameter accuracy of $\pm$1$\times $10$^{-3}$\,\AA.
The data treatment was made by Rietveld refinement with FULLPROF computer code in the assumption of  Thompson–Cox-Hastings line profile alloying multiple-phase refinement.
This  refinement
gives the weight
percentage of each of the coexisting phases, the line broadening leads to the autocoherent domain size owing to Scherrer formula.
The magnetic ordering temperatures $T_C$ were obtained on a differential sample magnetometer MANICS in a field of 1\,kOe on powder, in vacuum-sealed silica tube.
Curie temperature was determinate by extra-polating the linear part of the M–T curve and extending the baseline: intersection point corresponds to the value of T$_C$. The $M-H$ curves (Magnetic hysteresis measurement) were plotted at T= 10 and 293\,K with a Physical Properties Measurement System (PPMS) Quantum Design equipment and a maximum applied field of 90\,kOe on sample in epoxy resin.

\subsection{Structure analysis}
\begin{figure}
\centering
  \vskip -0.5 cm
    \includegraphics[height= 6.5 cm]{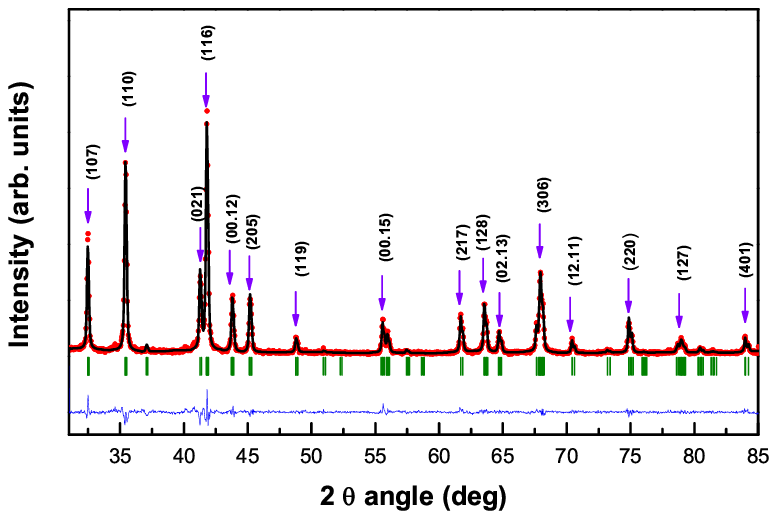} 
        \caption{\label{fig:1} Rietveld analysis for PrCo$_3$ ingot.} 
\vskip 0.5 cm
    \includegraphics[height = 6.5 cm]{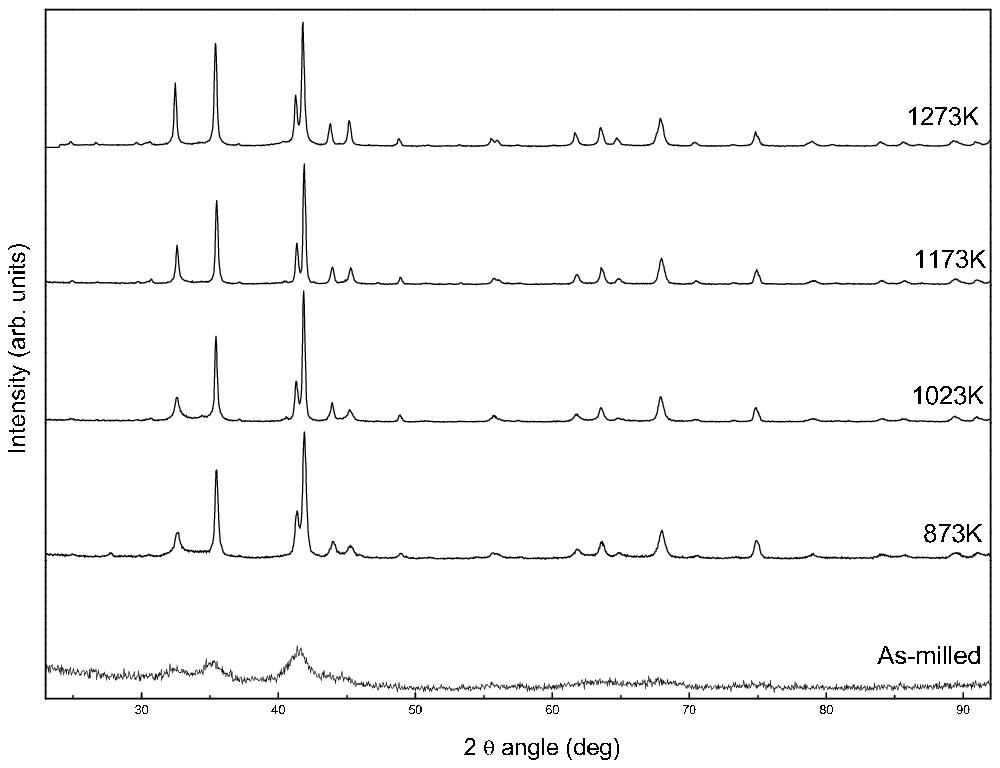} 
       \caption{\label{fig:2} XRD pattern of the PrCo$_3$ as-milled and annealed samples at 773 and 1273\,K during 30\,min.}

\label{D1}
\end{figure}

Before milling, the ingot is analyzed by X-ray diffraction (XRD). The Fig.~\ref{fig:1} presents the analysis result of XRD pattern of PrCo$_3$ ingot using the Rietveld analysis. The PrCo$_3$ ingot is nearly single phase with the rhombohedral PuNi$_3$-type structure (space group: $R\bar3m$) and is an appropriate precursor for using as pre-alloys to manufacture samples by high-energy milling.

In order to check the crystalline structure of PrCo$_3$ powders, the XRD measurement of samples annealed at 
distinct temperature have been performed. The XRD patterns of both as-quenched and annealed PrCo$_3$ powder 
are shown in Fig.\ref{fig:2}. It can be seen that as-quenched powder is essentially quasiamorphous. Such 
quasiamorphous matrix suggests the presence of various types of small crystallites. The diffraction peaks 
become more pronounced with increasing annealing temperature. The annealed samples at 773 and 1273\,K 
for 30\,min reveal the presence of PrCo$_3$ phases crystallized.

However, when the annealed is at temperature between 773 and 1023\,K, we observe in the low angles a large line around 32$^\circ$. The large peak widths points to a small grain size. When the powders are annealed at temperature higher than 1023\,K, the XRD patterns show the presence of PrCo$_3$ phases crystallized. We have observed a small amount (around 0.5\,wt$\%$) of Pr$_2$O$_3$ in all samples annealed after milling.
The mean diffraction crystallite size of the powder after annealing at 1023 and 1273\,K for 30\,min obtained by fitting the XRD diagram with Rietveld method, ranges between 31 and 45\,nm respectively, this implies that synthesized materials are noncrystalline.

The Curie temperature is determined by the exchange-coupling interaction between the nearest-neighbor Co-Co atoms, which is the origin of ferromagnetism. This interaction depends markedly on the interatomic distance.
The Curie temperature of PrCo$_3$ compound is 349\,K,~\cite{yrb10} this result is in agreement with the data given by Lemaire~\cite{l66}.

The low Curie temperature in PrCo$_3$ compounds is due to the short Co-Co interatomic distances at the sites $6c-18h$ and at $18h-18h$ ~\cite{yrb10}, where the Co atoms couple
antiparallel.
In this structure, the $6c-18h$ and at $18h-18h$ interactions are strongly negative.

\begin{figure}[h]
\begin{center}
\includegraphics[width = 0.6\textwidth]{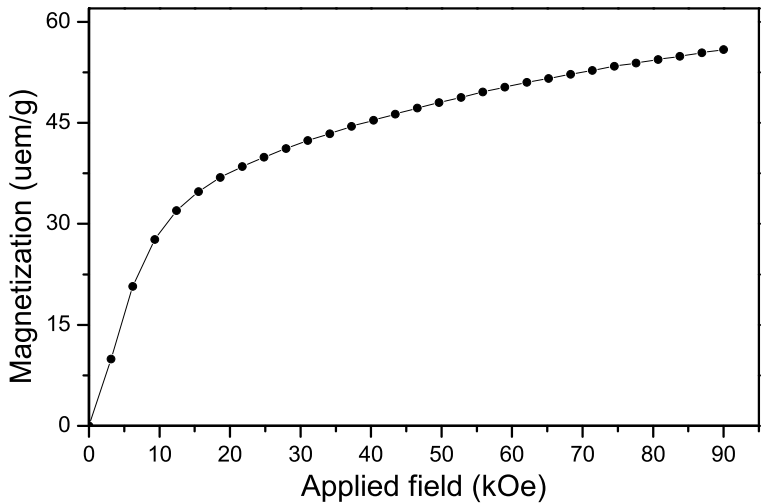}
\end{center}
\caption{\label{fig:3} Magnetization curves of PrCo$_3$ obtained after high-energy milling and annealed at 750$^\circ$C for 30\,min, measured at 293\,K.}
\end{figure}

Isothermal magnetization curves $M-H$ obtained at 293\,K is represented in Fig.~\ref{fig:3} for PrCo$_3$. This figure shows that at 293\,K the saturation is not reached, which gives evidence for a magnetocrystalline anisotropy among the highest known in those kinds of systems.
The pattern obtained on sample oriented under an external magnetic field, applied perpendicularly to the plane of the sample, shows only $(00l)$ Bragg peaks, which amount to suggest that the anisotropy strengthens the peaks of diffraction of type $(00l)$ and at room temperature the easy magnetization direction is parallel to the c-axis. A similar result is obtained for YCo$_{3-x}$Fe$_x$ by \cite{bgk00} obtained by oriented powder under extern field.

\subsection{Extrinsic magnetic properties}
\begin{figure}[h]
\begin{center}
\includegraphics[width = 0.9\textwidth]{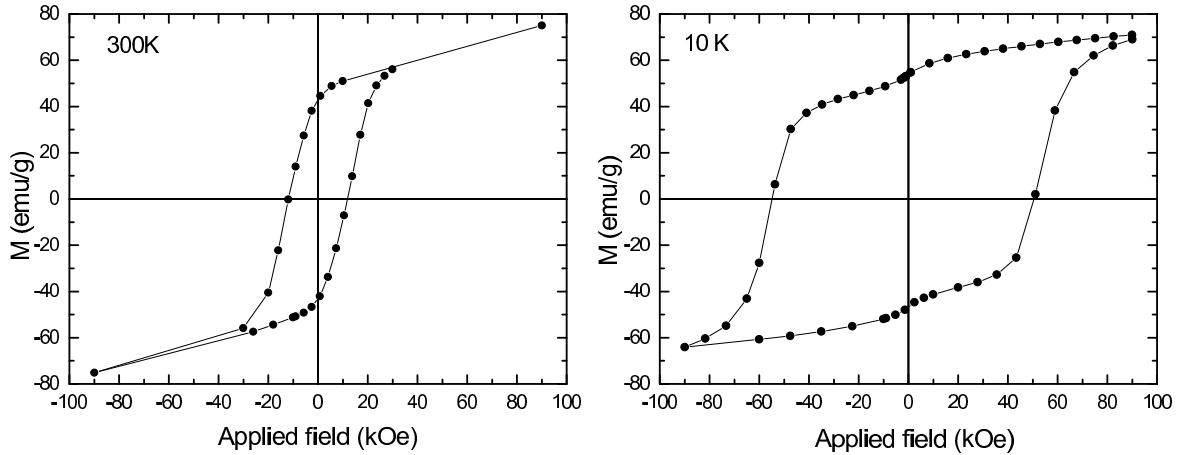}
\end{center}
\caption{\label{fig:4} Hysteresis loop of PrCo$_3$ milling before and after annealing at 1023\,K for 30\,min, measured at 293 and 10\,K.}
\end{figure}

The best magnetic properties were reached after a treatment at 1023~K for 30\,min. Fig.~\ref{fig:4} show as 
an example the hysteresis loops measured at 10 and 293\,K. This sample exhibits the highest coercivities
we have obtained, for $T= 10$\,K, $H_\mathrm{c}= 55$\,kOe, $M_\mathrm{R}= 56$\,emu/g and 
$M_\mathrm{R}/M_\mathrm{max}= 0.75$. For T= 293\,K, $H_\mathrm{c}= 12$\,kOe and $M_\mathrm{R}= 44$\,emu/g 
and $M_\mathrm{R}/M_\mathrm{max}= 0.60$ inherent to nanocrystalline state. The highest coercivity of 55\,kOe 
at 10\,K confirms the large magnetocrystalline anisotropy.

\section{Computational studies}
\subsection{Micromagnetism simulation}

In order to get a  picture of the link between the structure at the nanometer scale and the extrinsic magnetic properties we have
performed a micromagnetic simulation on a model for the system. The picture we have in mind is that the crystallization takes
place locally on dispersed seeds in the amorphous matrix representing the as-milled sample. Hence we consider an assembly of crystallites
presenting a uniaxial anisotropy comparable to what is expected in hard magnetic R-TM compounds. The saturation magnetization, $J_s$ of the
crystallites is that of the bulk PrCo$_3$, namely $J_s~=~0.723~T$. We consider the simplest realization of the model, by using identical
crystallites, located on the nodes of a simple cubic lattice. The crystallites are either spherical or cubic; however, the later case
is likely to be closer to the experimental sample. We have chosen somewhat arbitrarily a volumic fraction occupied by the crystallites
of $\phi~=~0.73$ (in the case of cubic crystallites) and the crystallite size corresponds to what has been estimated experimentally,
namely, a cubic edge $a~=~35~nm$. The amorphous character of the matrix is translated by a zero valued anisotropy constant, while its
saturation magnetization ($J_s^{(mat)}$) is a free parameter of the model. The value we get for $J_s^{(mat)}$, namely c.a.
$J_s/5$ seems reasonable in our picture of an amorphous matrix. The important point is that we clearly get the signature of
a two uncoupled phases behavior for $M(H)$; the amplitude of the kink at zero field being determined by $\phi$ and $J_s^{(mat)}$.

\begin{figure}[h]
\begin{center}
\includegraphics[width = 0.6\textwidth]{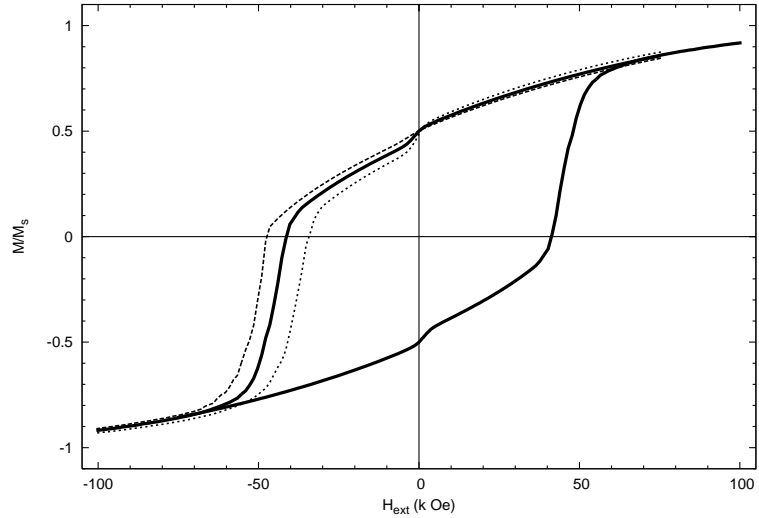}
\end{center}
\caption{\label{magpar_cub} Calculated de-magnetization curve in terms of the external field
for the systems of cubic inclusions with $\varphi= 0.73$.
Randomly distributed axes and matrix saturation magnetization
$J_s= 0.05$\,T, dashed line; $J_s= 0.125$\,T, solid line and $J_s= 0.25$\,T, dotted line.}
\end{figure}

\subsection{DFT calculations}
\subsubsection{Methodology}
Band structure spin polarized calculations were carried out by using the DFT scheme implemented in the projector-augmented wave (PAW) 
method, performed with the VASP package~\cite{Kre96,Kre99}. In the present study, the generalized gradient approximation (GGA) is 
used for the exchange and correlation energy functionals with the Perdew-Burke-Ernzerhof (PBE)~\cite{Per97} functional.
A plane wave basis set with a cutoff energy of 600\,eV was used in all calculations converged within 0.1\,eV in total energy. 
Brillouin zone sampling using a mesh of 0.5\,k points for each primitive cell of $R$Co$_x$ compounds, described respectively 
in $Fd\bar{3}m$ space group for $x$=2, $R\bar{3}m$ for $x$=3 and $P6_3/mmm$  for $x$=5 compounds. Structural optimization were 
carried out under the condition that residual forces should be smaller than 0.1\,eV.\AA$^{-1}$.
\subsubsection{Electronic structures}
\begin{figure}
\begin{center}
\includegraphics[width=10pc, angle =-90]{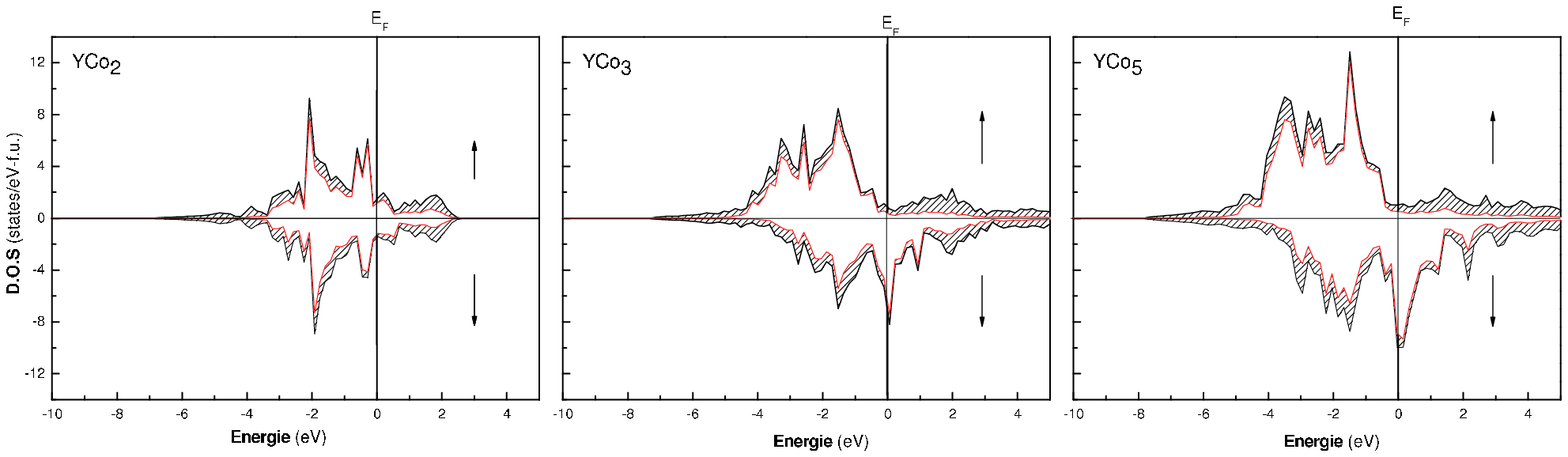}\\
\includegraphics[width=10pc, angle =-90]{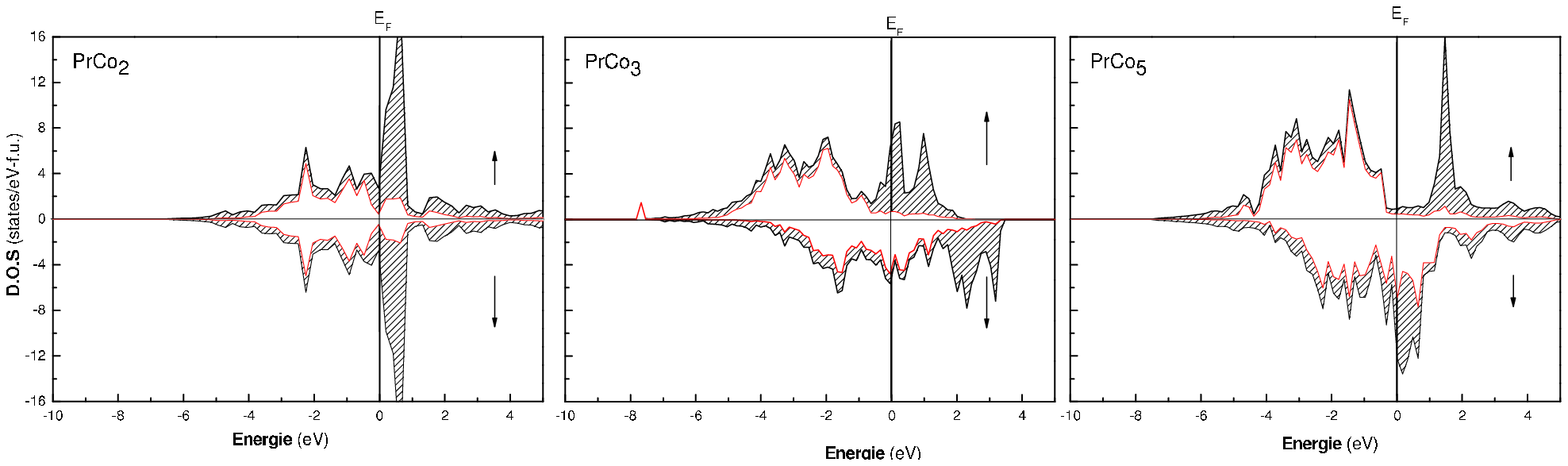}
\end{center}
\caption{\label{fig:DOS} Total density of states (shaded region) of $R$Co$_x$ ($R$=Y,Pr ; $x$=2, 3 and 5). The white area is corresponding 
to the only $3d$-states of Co.  The origin of the energy scale is located at the Fermi energy $E_F$.}
\end{figure}

Figure~\ref{fig:DOS} shows the density of states (DOS) for the YCo$_x$ and PrCo$_x$ ($x$=2, 3 and 5) in the two directions of spin.
For both series, the majority occupied states are dominated by the $3d$ states of Co, with a
small electronic charge transfert of $R$ to Co, which is more important from $R$=Y to Co than from Pr to Co (calculated by Bader's theory).
The increase of $x$ leads to a distinct shift of the Co-$3d$ states in spin majority and minority DOS as expected in the rigid-band Stoner theory. 
Whereas $R$Co$_2$ shows the paramagnetic state as the most stable in energy, the ferromagnetic behavior is confirmed for $x$=3 and $x$=5 compounds, 
with an increase of the angular magnetic moment from 1.45 to 7.34 and 3.59 to 4.70\,$\mu_B$\,f.u.$^{-1}$ for $R$=Y and Pr respectively. 
The magnetic anisotropy has been confirmed for $R$Co$_{3}$ compounds, since non-collinear spin calculation along the $z$~axis has 
been found $\sim$0.2\,eV\,f.u.$^{-1}$ more stable than calculation along $x$~axis direction. Since the DOS at Fermi level is high 
enough for $x$=3 and~5, the itinerant magnetism properties of theses compounds can be explained by intra-atomic $3d-3d$ exchange.
In contrast to YCo$_x$, PrCo$_x$ compounds present tiny and high density structures above the Fermi level, corresponding to localized 
empty $4f$ states of Pr. It is clear that the Co-$3d$ states are modified compared to YCo$_x$, due to the hybridization with the Pr-$4f$ states.
As expected in $3d-5d$ hybridization compounds, the energy between the bonding and anti-bonding sub-bands becomes different for the 
two-direction of spins~\cite{Ric98}. The consequence is a larger occupation of minority $5d$ states as compared to majority $5d$ of 
the rare earth, that means a ferrimagnetic coupling between $3d$ and $5d$ states occurs and increases with the increasing of $x$~composition.

\section{Conclusion and outlook}
The PrCo$_3$ intermetallic compound crystallizes in rhombohedral PuNi$_3$-type structure
and presents a very high uniaxial magnetocrystalline anisotropy. The coercivity of this compound,
prepared by high energy milling followed by annealing (1023~K, 30~mn), takes a very high value,
up to 55~kOe at 10~K. A micromagnetic simulation and atomic scale {\it ab-initio} calculations
have been performed to understand on the one hand the way in which the electronic structure
and especially the interplay between $3d$ and $4f$ electronic states,
leads to the observed anisotropy and magnetic moment, and on the other hand the relation between
the nanoscale structure and the extrinsic magnetic properties.
DFT calculations will be improved with spin-orbit coupling considerations, and with Hubbard corrections in order to better describe cohesive properties of materials with strong electronic correlations.
\section *{Acknowledgement}
The numerical part of this work was performed using HPC resources from GENSI- CINES (Grant  2010-096180).

\end{document}